\documentclass[journal,twocolumn]{IEEEtran}
\usepackage{caption}

\usepackage{cite}
\usepackage{multirow}
\usepackage{amsmath,amssymb,amsfonts}
\usepackage{algorithmic}
\usepackage{graphicx}
\usepackage{floatrow}
\usepackage{subcaption}
\usepackage{textcomp}
\usepackage{svg}
\usepackage{hyperref}
\usepackage[tableposition=top]{caption}
\usepackage{float}
\usepackage{xcolor}
\usepackage{mathtools, cuted}

\ifCLASSINFOpdf
\else
\fi

\begin{document}
%
\title{Availability Modeling for Blockchain Provisioning in Private Clouds}

\author{\IEEEauthorblockN{Jamilson Dantas\IEEEauthorrefmark{3}, Priscila Silva\IEEEauthorrefmark{2}, Lance Fiondella\IEEEauthorrefmark{2}, Carlos Melo\IEEEauthorrefmark{1}, and Paulo Maciel\IEEEauthorrefmark{3}}

\IEEEauthorblockA{\IEEEauthorrefmark{3}Centro de Informática (CIN), Universidade Federal de Pernambuco, Recife, Brazil \\
}
\IEEEauthorblockA{\IEEEauthorrefmark{2}Electrical and Computer Engineering, University of Massachusetts Dartmouth, MA, USA \\
}
\IEEEauthorblockA{\IEEEauthorrefmark{1}Centro de Ciências da Natureza (CCN), Universidade Federal do Piauí, Picos, Brazil \\
}

\IEEEauthorblockA{
carlos.alexandre@edu.ufpi.br\IEEEauthorrefmark{1},
\{psilva4, lfiondella\}@umaasd.edu\IEEEauthorrefmark{2},
\{prmm, jrd\}@cin.ufpe.br\IEEEauthorrefmark{3}
}
}

\maketitle

\begin{abstract}
Blockchain technology has emerged, and many previous studies have assessed its performance issues. However, less attention has been paid to the dependability attributes, which have been a critical topic in service provisioning, considering public or private infrastructures. This paper introduces analytical models to assess the availability of private blockchain infrastructure for Hyperledger Fabric-based applications. Furthermore, a case study will be presented to demonstrate the feasibility of the proposed model, which may assist stakeholders in deciding whether to migrate from old to new technology. Some of the obtained results indicate that, unlike most conventional systems, general availability may decrease as new nodes are added to the environment. This phenomenon occurs due to the adopted endorsement policy, which determines the proportion of required nodes to sign the authenticity of a transaction.
\end{abstract}


\IEEEpeerreviewmaketitle

\section{Introduction}

Blockchain is a decentralized technology that emerged in the late 2000s and is responsible for recording transactions across multiple computers securely and inflexibly, facilitating trust and transparency~\cite{kumaremploying}. Despite its foundational role in cryptocurrencies, blockchain remains relatively less known to users than the digital assets it supports, such as Bitcoin and non-fungible tokens (NFT) \cite{kumaremploying}, which may be attributed to the challenge of assigning value to an entire technology rather than to individual applications\cite{maciel2023performance}. However, particularly in permissioned networks that mostly focus on industrial applications, we need to explore its broader feasibility for adoption by organizations that do not share mutual trust. Permissioned networks provide enhanced security and control over data sharing and transactions in these environments; one of the most popular platforms for this mean is the Hyperledger Fabric (HLF) \cite{androulaki2018hyperledger}.



Previous studies~\cite{sukhwani2017performance,pongnumkul2017performance,thakkar2018performance} have examined various aspects of Hyperledger Fabric, primarily focusing on platform performance. However, some research efforts~\cite{melo2019blockchain}, have concentrated solely on system availability concerns, needing a comprehensive evaluation of the overall environment.

This paper's main contributions are as follows:
\begin{itemize}
    \item An overview of the Hyperledger Fabric environment and how it can be modeled;
    \item Development of generalized models for assessing the availability of Hyperledger Fabric environments;
    \item Presentation of a case study illustrating the practical application of the proposed models;
    \item Conducting simulations using the Mercury Modeling Tool to validate the effectiveness and feasibility of the proposed models.
\end{itemize}

The remainder of this paper is organized as follows. Section \ref{sec:rltd} presents the works that underlie this research. Section \ref{sec:hyperledger} provides an overview of the Hyperledger Fabric platform, its functioning, and how it can be modeled. Section \ref{sec:application} presents a high-level overview of an application running on a Hyperledger Fabric environment. Section \ref{sec:architectures} introduces the proposed models and their evaluation. Section \ref{sec:case} presents the scenarios used as a case study to demonstrate the feasibility of the proposed models. Finally, Section \ref{sec:conclusions} presents the final remarks and future directions.

\section{Related Works}
\label{sec:rltd}

The previous research on the Hyperledger Fabric platform has predominantly focused on performance metrics, leaving gaps in the exploration of dependability attributes. 
This paper addresses these gaps and provides an updated understanding of the platform's capabilities.

In \cite{melo_computing2022}, we introduced models for evaluating computational resource usage in Hyperledger Fabric (HLF) environments. 
Our study, which Leveraged Continuous Time Markov Chains (CTMCs) and stochastic Petri nets (SPNs), demonstrated the effectiveness of these formalisms in modeling HLF-based applications and detecting infrastructure bottlenecks. 
However, our current model focuses specifically on general availability and system uptime.

Other studies, such as \cite{jiang2020performance}, employed a hierarchical modeling approach to analyze performance metrics like throughput, latency, and system utilization. 
In contrast, \cite{wu_acm_ease2022,ke_springer_cc2022} developed a queue theory-based model focusing on HLF's transaction flow considering various service rates and their impact on the general performance.

Additionally, \cite{yuan2020performance} employed Generalized Stochastic Petri Nets (GSPN) to model the platform and investigate the influence of arrival rates and block sizes on HLF throughput and latency. 
While Sukhwani et al.\cite{sukhwani2018performance} used Stochastic Reward Networks (SRN). 
However, both models are isomorphic to the CTMCs presented in this paper, with a different focus.

It is worth noting that in both \cite{melo2019blockchain, melo2019models}, we evaluated availability and costs related to deploying blockchain applications in cloud computing environments. 
These works serve as the foundation for our current models, which form the basis of the framework proposed in this paper. 
Furthermore, our analysis extends beyond previous evaluations by considering endorsement policies, providing a comprehensive resource for blockchain service provisioning decisions.

\begin{table}[ht]
\centering
\begin{tabular}{|l|p{6.4cm}|}
\hline
\textbf{Paper} & \textbf{Main Differences and Contributions} \\ \hline
\cite{melo_computing2022} & Introduced models for evaluating computational resource usage in HLF environments using CTMCs and SPNs. Focused on general availability and system uptime. \\ \hline
\cite{jiang2020performance} & Employed a hierarchical modeling approach to analyze performance metrics such as throughput, latency, and system utilization. \\ \hline
\cite{wu_acm_ease2022,ke_springer_cc2022} & Developed a queue theory-based model focusing on HLF's transaction flow and considering various service rates' impact on general performance. \\ \hline
\cite{yuan2020performance} & Used GSPNs to model the platform and investigated the influence of arrival rates and block sizes on HLF throughput and latency. \\ \hline
\cite{sukhwani2018performance} & Used SRNs to model the platform's performance. \\ \hline
\cite{melo2019blockchain, melo2019models} & Evaluated availability and costs related to deploying blockchain applications in cloud computing environments. Considered endorsement policies, providing a comprehensive resource for blockchain service provisioning decisions. \\ \hline
This paper & Evaluates availability of blockchain applications in private environments with a generalizable model. \\ \hline
\end{tabular}
\caption{Summary of Differences and Contributions}
\label{tab:summary}
\end{table}

\section{A Hyperledger Fabric Overview}
\label{sec:hyperledger}

Hyperledger Fabric (HLF) is a platform for building and deploying solutions based on shared ledgers. Developed and maintained as an open-source standard by the Hyperledger Consortium under the auspices of the Linux Foundation, HLF offers a framework for creating distributed ledger applications. Various approaches exist for deploying a private or permissioned Hyperledger Fabric infrastructure. In a typical HLF environment, there are at least two key actors: the client and the service provider. The service provider often comprises a consortium of organizations that may not inherently trust each other. 

This paper focuses solely on evaluating the service provider side, with changes on the client side having no bearing on the overall system availability. On the service provider side, HLF environments leverage container technology and are centered around managing smart contracts, known as chaincodes. These chaincodes must be pre-installed during environment deployment and are responsible for enforcing business rules and orchestrating application activities. Conversely, the client side utilizes an SDK, typically implemented in Node.js or Java, to facilitate communication between the server and client components.

\begin{figure}[!ht]
\centering
\includegraphics[width=.75\textwidth]{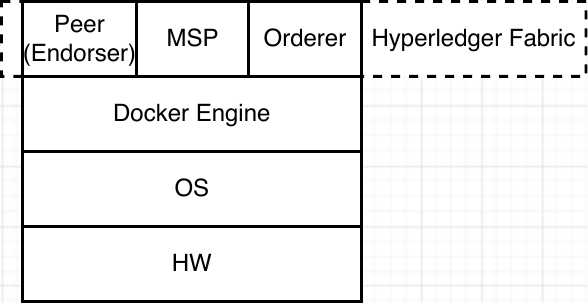}
\caption{Service Stack}
\label{fig:stack}
\end{figure}

Figure~\ref{fig:stack} provides a high-level depiction of a server hosting the minimal deployment of Hyperledger Fabric. As each component's failure leads to the service's failure, there are no explicit dependencies between them regarding service provisioning.

The high-level view encompasses four distinct components: the server's hardware (HW), the operating system (OS), the container engine (Docker Engine), and the deployed containers.
Within the Hyperledger Fabric environment, three types of containers are prominent:

\begin{itemize}
    \item Peer nodes, which execute endorsement tasks;
    \item MSP (Membership Service Provider) nodes manage membership and platform access;
    \item Orderer nodes receive transactions, group them into batches (blocks), and distribute them back to the peer nodes to persist it (commit).
\end{itemize}

\noindent This architectural overview forms the foundation for this paper's models based on the relationships among these components.

\section{A Blockchain-based Application}
\label{sec:application}

\noindent This paper evaluates a basic application deployed over the three Hyperledger Fabric's containers. Usually, a client uses its SDK to send a transaction to the service provider. However, many other steps are required and must be first accomplished on the other side in order for a transaction to be performed. 

\begin{figure}[ht]
    \centering
    \includegraphics[width=.99\textwidth]{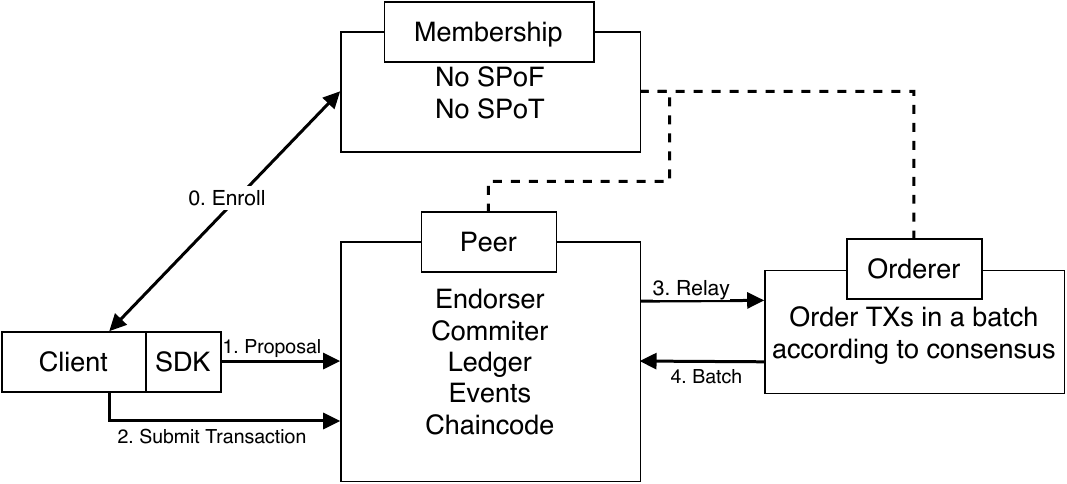}
    \caption{Hyperledger Fabric's Overview}
    \label{fig:overview}
\end{figure}

Figure~\ref{fig:overview} shows how the client and the service provider communicates, as well as which steps must be followed by a transaction in order for it to be fully accepted by the system \cite{hyperledgerintroduction,hyperledgerconsensus}.

The client connects to the system through the Membership Service Provider (MSP), which should provide no Single Point of Failure (SPoF) and no Single Point of Truth (SPoT). The MSP verifies the credentials of this client and allows access to the service provider. Later, the client proposes a transaction to the node known as Peer. There are many kinds of peers. We focus on endorsement peers, which simulate a transaction and send it back to the client with a signature that determines if the system can perform this transaction. As an example of a transaction, we may cite transferring assets between two clients, which requires both clients to exist and that the sender has as much balance as the value that he wants to transfer, discounting additional fees.



The transaction endorsement is based on a set of policies described by the chaincode. These policies specify how many and how these nodes agree with a transaction state. There are three main endorsement policies: AND, OR, and K-out-of-N (KooN). Suppose you have three servers on the service provisioning side, each hosting an endorsement peer. If an AND policy is used, all three nodes must endorse the transaction and sign it back to the client. If you use an OR policy, at least one of the three nodes must sign the transaction. The same applies to the KooN policy, where you determine how many available peers must sign the transaction, 1-out-of-3, 2-out-of-3, or even 3-out-of-3. After the endorsement, the transaction is submitted to the Orderer container. The orderer container joins all transactions in a batch (block) and sends it to the peers. Peers are responsible for executing transactions on the ledger following the requirements outlined in the chaincode, a fundamental aspect of Hyperledger Fabric applications.

\section{Availability Models}
\label{sec:architectures}
This section presents the proposed availability models that may represent environments that can host Hyperledger Fabric's nodes and blockchain-based applications.

We considered a two-stage hierarchical modeling to represent a Hyperledger Fabric-based architecture and the fact that all components must be operational to perform service provisioning. We adopted a Reliability Block Diagram (RBD) model in the first stage, representing the primary system's components. The RBD depicts the primary server components (hardware, operating system, and container engine (Docker Engine)).

\begin{figure}[ht]
 \centering
 \includegraphics[width=.9\textwidth]{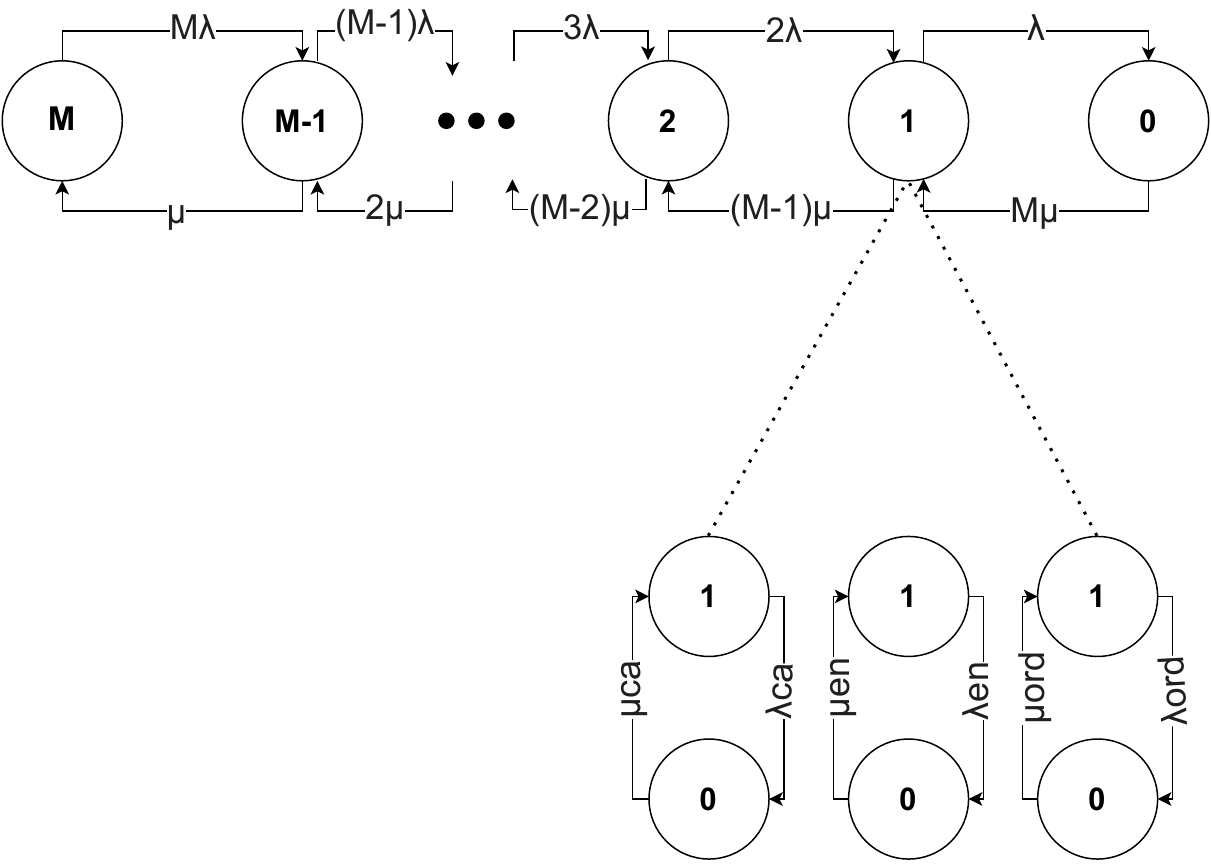}
 \caption{Availability CTMC Model}
 \label{fig:ctmc}
\end{figure}

The system's components are presented in a serial RBD, meaning that if at least one of its components fails, the whole system will also fail. After evaluating the primary component's RBD, we obtained their respective Mean Time to Failure (MTTF) and Mean Time to Repair (MTTR). The system's next modeling step deals with Hyperledger Fabric containers. Figure \ref{fig:ctmc} presents the Continuous Time Markov Chain (CTMC) model, which should conduct the second hierarchical modeling stage.

The CTMC first level deals with the previously stated RBD, which may have up to $M$ machines, each with its Hardware, Operating System, and Docker Engine. It is important to highlight that each machine in this model runs three containers (Endorser, MSP, and Orderer), represented by the CTMC's second level. The machines may enter a failure state or be repaired following an exponential distribution-based rate $\lambda$ and $\mu$, respectively. These rates are the inverse of the MTTF and MTTR values obtained from the RBD. The Hyperledger Fabric's containers already have their rates, which are the $\lambda_{ca}$ and $\mu_{ca}$ that stands for the MSP fail. Repair rates, $\lambda_{en}$ and $\mu_{en}$ for the endorser container, and $\lambda_{ord}$ and $\mu_{ord}$ represents the orderer rates.

At this point, we could formulate a set of equations to evaluate the system's availability based on the aimed endorsement policy by using the State Diagrams package in Wolfram Mathematica~\cite{weisstein2004mathworld} and concepts presented by~\cite{maciel2023performance}. Equation~\ref{eq:server} depicts a model with only one physical machine and its three containers, which enable availability (A) computations of a single server. Later, we generalized this equation using a binomial distribution.


Equation~\ref{eq:koon} denotes the K-out-of-N endorsement policy, where someone can establish a $K$ number of an $M$ total of components that must be operational to accomplish service provisioning, which means the system is available and the endorsement is performed. This expression is a generalization of the previous one and can be used to calculate any number of servers and their associated container.

\begin{equation}
\begin{split}
    A_{\text{Server}} = \left ( \frac{\mu}{\mu + \lambda }  \right ) \times \left ( \frac{\mu_{ca}}{\mu_{ca} + \lambda_{ca} }  \right ) \times \left ( \frac{\mu_{en}}{\mu_{en} + \lambda_{en} }  \right ) \times \\ \times \left ( \frac{\mu_{ord}}{\mu_{ord} + \lambda_{ord} }  \right )
    \label{eq:server}
\end{split}
\end{equation}


\begin{equation}
    A_{\text{KooN}} = \sum_{i=k}^{M}\binom{M}{k} A_{\text{Server}}^{k}(1-A_{\text{Server}})^{M-k}
    \label{eq:koon}
\end{equation}

\noindent where $K$ stands for the number of components expected to be operational, and $M$ is the total of resources that we have. Some specific scenarios may be extracted from the KooN policy, meaning that some other expressions can be obtained to calculate a combination of $K$ and $N$ values. The third Expression \ref{eq:and} represents the AND endorsement policy, which requires that all components in the first and second levels of the CTMC be operational, which means that it is an $N$-out-of-$N$.

\begin{equation}
\begin{split}
    A_{\text{NooN}} = \left ( \frac{\mu}{\mu + \lambda }  \right )^{M} \times \left ( \frac{\mu_{ca}}{\mu_{ca} + \lambda_{ca} }  \right )^{M} \times \left ( \frac{\mu_{en}}{\mu_{en} + \lambda_{en} }  \right )^{M} \times \\ \times \left ( \frac{\mu_{ord}}{\mu_{ord} + \lambda_{ord} }  \right )^{M}
    \label{eq:and}
\end{split}
\end{equation}



\section{Results and discussion}
\label{sec:case}

This section provides a case study that demonstrates the proposed model's feasibility. We are considering an availability evaluation scenario. 

The initial action involves gathering the necessary input values from the system to input into the models and conducting the availability assessment. These values comprise a mix of sources: some were sourced from relevant literature \cite{dantas2013,melo2016,sebastio2018availability}, while others were drawn from manufacturer charts and white papers. Table \ref{tab:input_baseline} presents the input values utilized in this study. Table~\ref{tab:scenarios} shows ten scenarios employed for illustration, where each considers a possible combination of endorsement policy and several nodes varying from one up to four servers hosting the three containers.


\begin{table}[htpb]
\centering
\caption{Input Parameters for Availability Evaluation}
\label{tab:input_baseline}

\begin{tabular}{l|r|r}
\hline
\multicolumn{1}{c|}{\textbf{Component}}   & \multicolumn{1}{c|}{\textbf{MTTF (h)}} & \multicolumn{1}{c}{\textbf{MTTR (h)}} \\ 
\hline
Hardware (HW)  & 8760 & 1.66 \\
Operating System (OS) & 2893 & 0.15 \\
Docker Engine (DE) & 2516 & 0.15 \\
Containers & 1258 & 0.15 \\ \hline
\end{tabular}
\end{table}


\begin{table}[htpb]
\caption{Evaluated Scenarios}
\label{tab:scenarios}
\begin{tabular}{r|c|c|c}
\hline
\multicolumn{1}{c|}{\textbf{Scenario}} & \multicolumn{1}{c|}{\textbf{Policy}} & \multicolumn{1}{c|}{\textbf{Required Servers}} & \multicolumn{1}{c}{\textbf{Total Servers}} \\ 
\hline
\hline
1                            & AND                        & 1                                    & 1                                 \\
2                            & AND                        & 2                                    & 2                                 \\
3                            & AND                        & 3                                    & 3                                 \\
4                            & AND                        & 4                                    & 4                                 \\
5                            & OR                         & 1                                    & 2                                 \\
6                            & OR                         & 1                                    & 3                                 \\
7                            & OR                         & 1                                    & 4                                 \\
8                            & KooN                       & 2                                    & 3                                 \\
9                            & KooN                       & 2                                    & 4                                 \\
10                           & KooN                       & 3                                    & 4                                 \\ \hline
\end{tabular}
\end{table}


After enumerating input availability values and exploring ten distinct scenarios following OR, AND, or K-out-of-N endorsement policy, we assessed their respective availabilities using Equation~\ref{eq:koon}. Table \ref{tab:av_results} showcases the resultant general availability for each scenario proposed.

\begin{table}[htpb]
\caption{Availability Results}
\label{tab:av_results}
\begin{tabular}{r|r|r|r}
\hline
\multicolumn{1}{l|}{\textbf{Scenario}} & \multicolumn{1}{l|}{\textbf{Av. (\%)}} & \multicolumn{1}{l|}{\textbf{Av. (\#9s)}} & \multicolumn{1}{l}{\textbf{A. Downtime (h)}} \\ 
\hline
\hline
1  & 99.9341 & 3.18     & 5.77    \\
2  & 99.8683 & 2.88     & 11.53   \\
3  & 99.8026 & 2.70     & 17.29   \\
4 & 99.7369 & 2.58     & 23.05   \\
5 & 99.9998 & 5.70     & 0.0038  \\
6  & 99.9999 & 9.45     & 0.000003\\
7  & 99.9999 & 12.66    & 0.00000002\\
8 & 99.9987 & 4.89     & 0.011   \\
9 & 99.9999 & 8.90     & 0.000009\\
10 & 99.9997& 5.52     & 0.022   \\ \hline
\end{tabular}
\end{table}

The availability of an AND policy decreases with the addition of more resources (nodes), as expected, due to the necessity for all components to remain operational. Failure of any component prevents transaction endorsement. Conversely, an OR endorsement policy experiences increased availability with the addition of more resources. The KooN endorsement policy demonstrates intermediate availability compared to AND and OR policies. Its availability diminishes with the requirement for more nodes, resembling an AND policy. This outcome aligns with the Annual Downtime analysis for each scenario and endorsement policy.

We have experimented to understand the impact of the input parameters on the overall system's annual downtime. To accomplish this task, we varied and rounded each parameter value in +50\% and -50\% as in the given time interval.

\begin{figure*}
    \centering
    
    \begin{subfigure}[b]{0.42\textwidth}
    \centering
    \includegraphics[width=\textwidth]{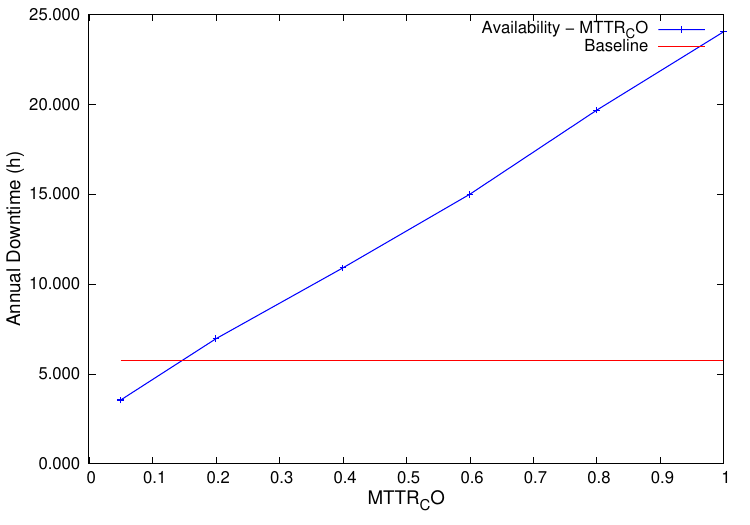}
    \caption{Annual Downtime (h) x $MTTR\_CO$}
    \label{fig:mttrco}
    \end{subfigure}%
    ~ 
    \begin{subfigure}[b]{0.42\textwidth}
    \centering                \includegraphics[width=\textwidth]{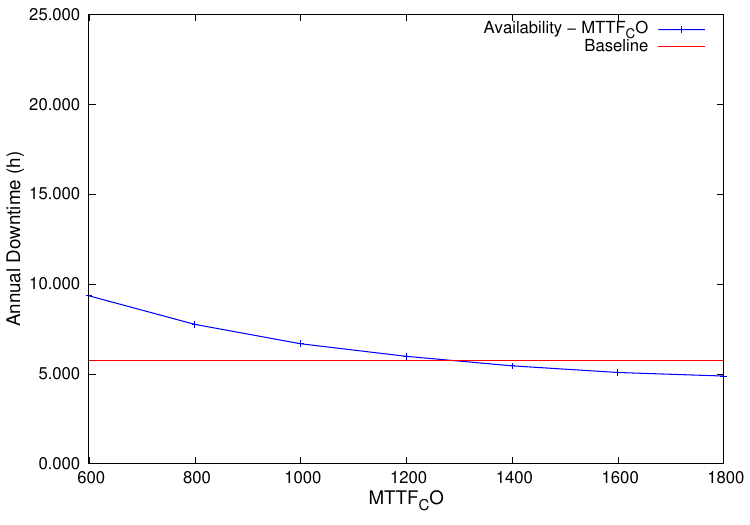}    
    \caption{Annual Downtime (h) x $MTTF\_CO$}
    \label{fig:mttfco}
    \end{subfigure}%
    
    \begin{subfigure}[b]{0.42\textwidth}
    \centering                \includegraphics[width=\textwidth]{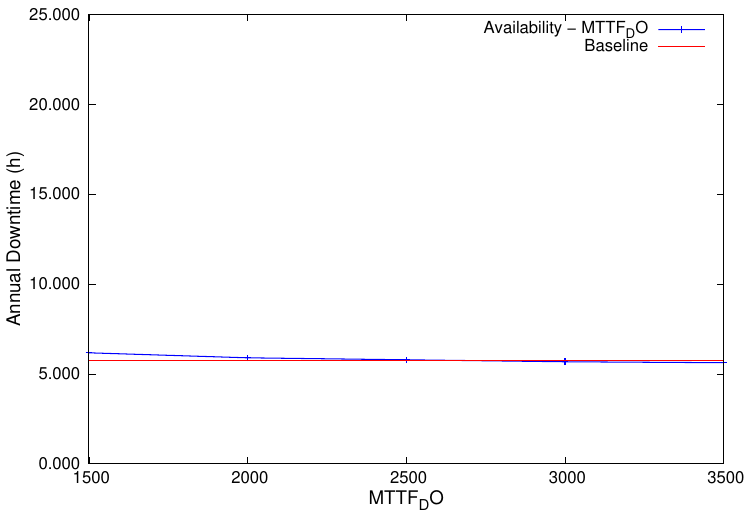}
    \caption{Annual Downtime (h) x $MTTF\_DO$}
    \label{fig:mttfdo}
    \end{subfigure}
    ~
    \begin{subfigure}[b]{0.42\textwidth}
    \centering                \includegraphics[width=\textwidth]{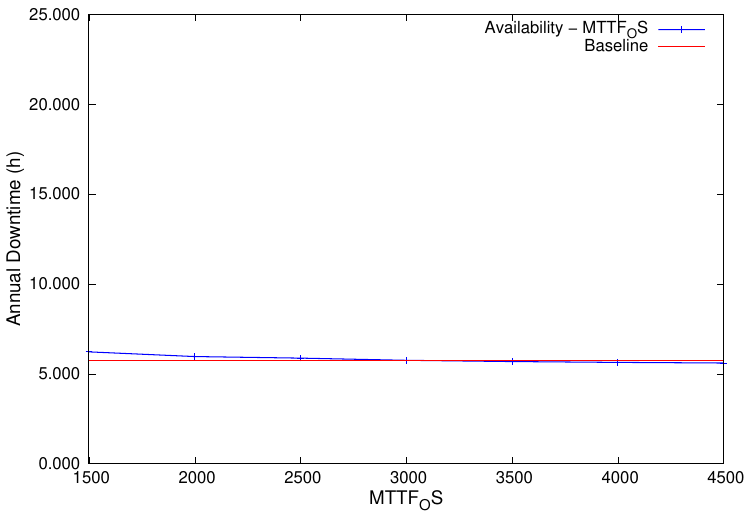}
    \caption{Annual Downtime (h) x $MTTF\_OS$}
    \label{fig:mttfos}
    \end{subfigure}
    
    \caption{Sensitivity analysis for Availability}
    \label{fig:sensitivity-ava}
\end{figure*}

The bottleneck findings are illustrated in Figure \ref{fig:sensitivity-ava}, where we highlight the parameters exerting the greatest and least influence on system availability. We can see from this figure that the higher the MTTF, the lower the annual downtime, while the lower the MTTR, the higher the obtained downtime values, which should be expected behavior. Among these results, we may highlight the Container's MTTR and MTTF (\ref{fig:mttrco} and \ref{fig:mttfco} subfigures). Those parameters are the bottleneck regarding the system's availability-associated metrics. On the other hand, the Docker MTTF and the Operating System MTTF do not mean to impact the interesting metric (see Figures \ref{fig:mttfdo} and \ref{fig:mttfos}).

\section{Conclusions and Future Works}
\label{sec:conclusions}

This paper presented a set of hierarchical models to evaluate the availability of the Hyperledger Fabric platform within private cloud computing infrastructures. The results may help stakeholders and decision-makers assess the impact of varying the parameters on a generalizable model representing the HLF blockchain environment to determine the effects on general availability and associated annual downtime. We presented a case study that dealt with the availability evaluation of a set of scenarios varying the number of nodes responsible for endorsement, an important characteristic of a Hyperledger Fabric infrastructure. 

The main limitation of the current work is the need for an experimental evaluation to validate the proposed model in a real environment. This means that the obtained results are simulated based on the service provisioning stack and related work's provided data. Continuously, we intend to evaluate and compare different scenarios and applications, such as Industry 4.0, disaster risk management, and security systems.

\bibliographystyle{IEEEtran}
\bibliography{sample}
\end{document}